\documentclass[11pt, a4paper]{article}
\usepackage[textwidth = 7in, textheight=25cm,nohead]{geometry}
\usepackage{ulem}
\usepackage{arydshln}
\usepackage{graphicx, amssymb, amsmath, amsthm, amstext, booktabs, float, multirow, rotating,natbib,bm}
\usepackage{caption}
\usepackage{subcaption}
\usepackage{stackengine}
\usepackage{amssymb}
\usepackage{mathtools,bm}
\usepackage[mathscr]{euscript}

\usepackage{bbding}
\usepackage{pifont}

\usepackage{xparse}

\usepackage{mathdots}

\usepackage{tikz}
\usepackage{setspace}
\usepackage{enumerate}
\usepackage{multirow}
\usepackage{graphicx,nicefrac}

\numberwithin{figure}{section} 
\numberwithin{table}{section}

\newtheorem{theorem}{Theorem}[]

\newtheorem{frenchthm}{Th\'eor\`eme}[]
\newtheorem{frenchdef}{D\'efinition}[]

\theoremstyle{remark}

\theoremstyle{definition}

\theoremstyle{definition}

\bibpunct{(}{)}{,}{a}{,}{,}

\makeatletter
\newcommand{\doublewidetilde}[1]{{%
		\mathpalette\double@widetilde{#1}%
}}
\newcommand{\double@widetilde}[2]{%
	\sbox\z@{$\m@th#1\widetilde{#2}$}%
	\ht\z@=.9\ht\z@
	\widetilde{\box\z@}%
}
\makeatother

 \let\mathscr\relax% just so we can load this and rsfs
\usepackage[scr]{rsfso}

\def\ppn{\vskip 6pt \noindent }
\def\R{{\mathbb{R}}}
\def\N{{\mathbb{N}}}

\newcommand{{\Xs}}{{\cal X}}
\newcommand{{\Xss}}{{\mathfrak{X}}}
\newcommand{{\Ys}}{{\cal Y}}
\newcommand{{\Yss}}{{\mathfrak{Y}}}
\newcommand{{\Ls}}{{\cal L}}
\newcommand{{\Ss}}{{\cal S}}
\newcommand{{\Ms}}{{\cal M}}
\newcommand{{\Gs}}{{\cal G}}
\newcommand{{\Hs}}{{\cal H}}
\newcommand{{\Ns}}{{\cal N}}
\newcommand{{\Is}}{{\cal I}}
\newcommand{{\Vs}}{{\cal V}}
\newcommand{{\Ds}}{{\cal D}}
\newcommand{{\As}}{{\cal A}}
\newcommand{{\Bs}}{{\cal B}}
\newcommand{{\Cs}}{{\cal C}}
\newcommand{{\Rs}}{{\cal R}}
\newcommand{{\Es}}{{\cal E}}
\newcommand{{\Fs}}{{\cal F}}
\newcommand{{\Us}}{{\cal U}}
\newcommand{{\Ps}}{{\cal P}}
\newcommand{{\ttheta}}{{\bm{\theta}}}
\newcommand{{\Ttheta}}{{\bm{\Theta}}}
\newcommand{{\Oomega}}{{\bm{\Omega}}}
\newcommand{{\oomega}}{{\bm{\omega}}}
\newcommand{{\mmu}}{{\bm{\mu}}}
\newcommand{{\ggamma}}{{\bm{\gamma}}}
\newcommand{{\Ssigma}}{{\bm{\Sigma}}}
\newcommand{{\Sss}}{{\bm{\Ss}}}
\newcommand{{\pp}}{{\mathbf p}}
\newcommand{{\ww}}{{\mathbf w}}
\newcommand{{\mm}}{{\mathbf m}}
\newcommand{{\bb}}{{\mathbf b}}
\newcommand{{\uu}}{{\mathbf u}}
\newcommand{{\ppi}}{{\bm{\pi}}}
\newcommand{{\phhi}}{{\bm{\phi}}}
\newcommand{{\pssi}}{{\bm{\psi}}}
\newcommand{{\XX}}{{\mathbf X}}
\newcommand{{\UU}}{{\mathbf U}}
\newcommand{{\BB}}{{\mathbf B}}
\newcommand{{\KK}}{{\mathbf K}}
\newcommand{{\HH}}{{\mathbf H}}
\newcommand{{\II}}{{\mathbf I}}
\newcommand{{\PP}}{{\mathbf P}}
\newcommand{{\yy}}{{\mathbf y}}
\newcommand{{\ee}}{{\mathbf e}}
\newcommand{{\ab}}{{\mathbf a}}

\newcommand{{\dd}}{{\mathbf d}}
\newcommand{{\zero}}{{\mathbf 0}}
\newcommand{{\uno}}{{\mathbf 1}}

\newcommand{{\DDelta}}{{\bm \Delta}}
\newcommand{{\piXY}}{{\pi_{{\scriptscriptstyle XY}}}}
\newcommand{{\FXY}}{{F_{{\scriptscriptstyle XY}}}}
\newcommand{{\piot}}{{\pi_{{\scriptscriptstyle 12}}}}
\newcommand{{\piXYstar}}{{\pi^*_{{\scriptscriptstyle XY}}}}
\newcommand{{\piXYstarstar}}{{\pi^{**}_{{\scriptscriptstyle XY}}}}
\newcommand{{\piotstar}}{{\pi^*_{{\scriptscriptstyle 12}}}}
\newcommand{{\piX}}{{\pi_{{\scriptscriptstyle X}}}}
\newcommand{{\FX}}{{F_{{\scriptscriptstyle X}}}}
\newcommand{{\pio}}{{\pi_{{\scriptscriptstyle 1}}}}
\newcommand{{\piXstar}}{{\pi^*_{{\scriptscriptstyle X}}}}
\newcommand{{\piostar}}{{\pi^*_{{\scriptscriptstyle 1}}}}
\newcommand{{\piY}}{{\pi_{{\scriptscriptstyle Y}}}}
\newcommand{{\FY}}{{F_{{\scriptscriptstyle Y}}}}
\newcommand{{\pit}}{{\pi_{{\scriptscriptstyle 2}}}}
\newcommand{{\piYstar}}{{\pi^*_{{\scriptscriptstyle Y}}}}
\newcommand{{\pitstar}}{{\pi^*_{{\scriptscriptstyle 2}}}}
\newcommand{{\phiXY}}{{\varphi_{{\scriptscriptstyle XY}}}}
\newcommand{{\phiot}}{{\varphi_{{\scriptscriptstyle 12}}}}
\newcommand{{\phiXYstar}}{{\varphi^*_{{\scriptscriptstyle XY}}}}
\newcommand{{\phiotstar}}{{\varphi^*_{{\scriptscriptstyle 12}}}}
\newcommand{{\phiX}}{{\varphi_{{\scriptscriptstyle X}}}}
\newcommand{{\phio}}{{\varphi_{{\scriptscriptstyle 1}}}}
\newcommand{{\phiXstar}}{{\varphi^*_{{\scriptscriptstyle X}}}}
\newcommand{{\phiostar}}{{\varphi^*_{{\scriptscriptstyle 1}}}}
\newcommand{{\phiYstar}}{{\varphi^*_{{\scriptscriptstyle Y}}}}
\newcommand{{\phitstar}}{{\varphi^*_{{\scriptscriptstyle 2}}}}
\newcommand{{\phiY}}{{\varphi_{{\scriptscriptstyle Y}}}}
\newcommand{{\phit}}{{\varphi_{{\scriptscriptstyle 2}}}}
\newcommand{{\phiYgX}}{{\varphi_{{\scriptscriptstyle Y|X}}}}
\newcommand{{\phitgo}}{{\varphi_{{\scriptscriptstyle 2|1}}}}
\newcommand{{\phiXgY}}{{\varphi_{{\scriptscriptstyle X|Y}}}}
\newcommand{{\phiogt}}{{\varphi_{{\scriptscriptstyle 1|2}}}}
\newcommand{{\phiXgYstar}}{{\varphi^*_{{\scriptscriptstyle X|Y}}}}
\newcommand{{\phiogtstar}}{{\varphi^*_{{\scriptscriptstyle 1|2}}}}
\newcommand{{\phiYgXstar}}{{\varphi^*_{{\scriptscriptstyle Y|X}}}}
\newcommand{{\phitgostar}}{{\varphi^*_{{\scriptscriptstyle 2|1}}}}

\newcommand{{\toL}}{{\overset{\mathcal{L}}{\longrightarrow}\ }}

\newcommand{{\MC}}{{\,  *_{\text{\scalebox{0.65}{$\Ms$}}}\,  }}

\newcommand{{\dou}}{$\leadsto$\ }

\DeclareMathOperator{\Ran}{Ran}

\begin{document}
	
	\setlength{\belowdisplayskip}{5pt} \setlength{\belowdisplayshortskip}{3pt}
	\setlength{\abovedisplayskip}{5pt} \setlength{\abovedisplayshortskip}{0pt}
	
	\title{(Re-)reading Sklar (1959) -- A personal view on Sklar's theorem}
	
	\author{\sc{Gery Geenens}\thanks{{\tt ggeenens@unsw.edu.au}}\\School of Mathematics and Statistics,\\ UNSW Sydney, Australia 
	}
	
	\date{}
	\maketitle
	\thispagestyle{empty} 
	
	\ppn In probability and statistics, copula methods have become ubiquitous when it comes to analyse, model and quantify the dependence between variables. Systematically, any written (research paper) or verbal (conference talk) communication about copulas starts with a statement of so-called `{\it Sklar's theorem}', establishing existence of a copula for any multivariate probability distribution. After defining a $d$-dimensional copula ($d \in \N$) as a continuous cumulative distribution function supported on the unit hypercube $[0,1]^d$ with uniform marginals, the theorem is typically stated under a form equivalent to the following: 
	
	\begin{theorem}[{\bf `Sklar's theorem'}] \label{thm:pretendSklar} 
		\begin{enumerate}[a)]
			\item Let $F_{1\ldots d}$ be a $d$-dimensional ($d \in \N$) distribution function with marginals $F_1,\ldots,F_d$. Then, $(i)$ there exists a $d$-dimensional copula $C$ such that \begin{equation} F_{1\ldots d}(x_1,\ldots,x_d) = C(F_1(x_1),\ldots,F_d(x_d)) \label{eqn:Sklar} \tag{$\star$}\end{equation}
			for all $(x_1,\ldots,x_d) \in \overline{\R}^d$; $(ii)$ if each $F_k$ ($k=1,\ldots,d$) is continuous, then $C$ is unique; otherwise, $C$ is uniquely determined on $\bigtimes_{k=1}^d \Ran F_k$, where $\Ran F_k = \{t \in [0,1]: \exists x \in \overline{\R} \text{ s.t. } F_k(x) = t\}$.
			\item Conversely, if $C$ is a $d$-dimensional copula and $F_1,\ldots,F_d$ are univariate distribution functions, then the function $F_{1\ldots d}$ defined via (\ref{eqn:Sklar}) is a $d$-dimensional distribution function with marginals $F_1,\ldots,F_d$.
		\end{enumerate}
	\end{theorem}

\ppn (Here $\overline{\R}$ denotes the extended real line $[-\infty,\infty]$.) This is the theorem as it is stated in \citet[Theorem 5.3]{McNeil05} and (for $d=2$) in \citet[Theorem 2.3.3]{Nelsen06}. Statements in other main references on copulas, such as \citet[Theorem 1.1]{Joe2015}, \citet[Theorem 2.2.1]{Durante15} or \citet[Theorem 2.3.1]{Hofert18}, differ only slightly. The reference provided is invariably \cite{Sklar59}.

\ppn Now, not long ago, in the discussion which followed a seminar on copulas which I attended, the speaker argued that \cite{Sklar59} was certainly the most cited {\it unread} statistical paper. The argument holds water if we put in perspective the facts that $(i)$ \cite{Sklar59} is referenced each time copulas are introduced, leading to a huge number of citations (close to 11,000 at the time of writing this note, according to {\it Google Scholar}); and $(ii)$ it is an `old' paper in French, which was difficult to access for a long time, even after it was republished \citep{Bosq10}. Thus, as Theorem \ref{thm:pretendSklar} above is found (in English) in a multitude of other sources easily accessible, it may be reasonably conjectured that only a minor fraction of the authors citing \cite{Sklar59} put in the effort to access and read the original text. 

\ppn Admittedly, I was not part of that minor fraction until recently; and Theorem \ref{thm:pretendSklar} was reported as-is in \cite{Geenens2017} and \cite{Geenens2020} with reference to \cite{Sklar59} (which is very bad practice, for that matter). Yet, the previous discussion prompted me to read the original paper in French, only to find out that \cite{Sklar59} does {\it not} contain any such `Sklar's theorem' under the above form -- making a wider community aware of this fact may be the only purpose of this short note. In fact, the paper comprises five theorems, among which three ({\it Th\'eor\`eme 1}, {\it Th\'eor\`eme 2} and {\it Th\'eor\`eme 3}), when combined, allow one to reconstruct and/or deduce Theorem \ref{thm:pretendSklar}. For convenience, we translate\footnote{Bearing in mind that `{\it traduire c'est trahir}' -- `translating is betraying' -- as my high-school English teacher used to say. Ironically, the statements in \cite{Sklar59} were themselves, presumably, French translations of Sklar's initial thoughts, making all this an interesting instance of the `broken telephone game'.}  here in English \cite{Sklar59}'s {\it Th\'eor\`eme 1}, {\it Th\'eor\`eme 2} and {\it Th\'eor\`eme 3}, as well as the definition of a copula appearing in the sequence ({\it D\'efinition 1}). (The notations and footnotes are original from \cite{Sklar59}. The three {\it Th\'eor\`emes} and the {\it D\'efinition} appear in this order. Nothing is omitted between the statements, given without proofs.)

\noindent\rule{\textwidth}{1pt}

	\begin{frenchthm} \label{thm:Sklar1} Let $G_n$ be an $n$-dimensional cumulative distribution function with margins $F_1,F_2,\ldots,F_n$. Let $R_k$ be the set of values of $F_k$, for $k=1,\ldots,n$. Then there exists a unique function $H_n$ defined on the Cartesian product $R_1 \times R_2 \times \ldots R_n$ and such that 
		\[G_n(x_1,\ldots,x_n) = H_n(F_1(x_1),F_2(x_2),\ldots,F_n(x_n)). \]
	\end{frenchthm}
	\begin{frenchdef} \label{def:copsklar} We call ($n$-dimensional) \uline{copula} any function $C_n$, continuous, non-decreasing\footnote{In the sense of an $n$-dimensional cumulative distribution function.}, defined on the Cartesian product of $n$ closed intervals $[0,1]$ and satisfying the conditions:
		\[C_n(0,0,\ldots,0) = 0 , \quad C_n(1,\ldots,1,\alpha,1,\ldots,1) = \alpha. \footnote{Special cases of such functions were considered in \cite{Feron56b}.} \] 
	\end{frenchdef}
	\begin{frenchthm} \label{thm:Sklar2} The function $H_n$ of Theorem \ref{thm:Sklar1} can be extended (in general, in more than one way) into a copula $C_n$. An extension of $H_n$, the copula $C_n$ satisfies the condition:
		\[G_n(x_1,\ldots,x_n) = C_n(F_1(x_1),F_2(x_2),\ldots,F_n(x_n)). \]
	\end{frenchthm}
	
	\newpage
	
	\begin{frenchthm} \label{thm:Sklar3} Let be given univariate cumulative distribution functions $F_1, F_2,\ldots,F_n$. Let $C_n$ be an arbitrary $n$-dimensional copula. Then the function $G_n$ defined as 
		\[G_n(x_1,\ldots,x_n) = C_n(F_1(x_1),F_2(x_2),\ldots,F_n(x_n))\]
		is an $n$-dimensional cumulative distribution function with margins $F_1, F_2,\ldots,F_n$.
	\end{frenchthm}
	
\noindent\rule{\textwidth}{1pt}

\ppn {\it Th\'eor\`eme} \ref{thm:Sklar1} establishes the {existence} and {uniqueness} of the function that would later be called {\it subcopula} -- \cite{Sklar59} did not use that word, not introduced before \citet[Definition 3]{Schweizer74}. This subcopula, denoted $H$ below to stay consistent with {\it Th\'eor\`eme} \ref{thm:Sklar1}, is defined only on the `Cartesian product of the sets of values of $F_k$', that is, $\bigtimes_{k=1}^d \Ran F_k$ in the notation of Theorem \ref{thm:pretendSklar}, and satisfies
\begin{equation} F_{1\ldots d}(x_1,\ldots,x_d) = H(F_1(x_1),\ldots,F_d(x_d)) \qquad  \forall (x_1,\ldots,x_d) \in \overline{\R}^d.  \label{eqn:Sklarsub} \tag{$\star\star$}\end{equation}
%We deduce that, for $(u_1,\ldots,u_d) \in \bigtimes_{k=1}^d \Ran F_k$,
%\[ S(u_1,\ldots,u_d) = F_{1\ldots d}(F_1^{(-1)}(u_1),\ldots,F_d^{(-1)}(u_d))\]
%where $F_k^{(-1)}$ is the quantile function of $F_k$ (i.e., the generalised inverse of $F_k$, $F_k^{(-1)}(u_k) = \inf\{x \in \overline{\R}: F_k(x) \geq u_k\}$). 
Although no details are given, {\it Th\'eor\`eme} \ref{thm:Sklar2} states that the unique subcopula satisfying  (\ref{eqn:Sklarsub}) may be extended `{\it in general, in more than one way}' beyond $\bigtimes_{k=1}^d \Ran F_k$ into a function defined on the whole of the unit hypercube $[0,1]^d$ and satisfying {\it D\'efinition} \ref{def:copsklar}, such a function being called {\it copula}. In other words, there exists at least one copula $C$ coinciding exactly with the subcopula on $\bigtimes_{k=1}^d \Ran F_k$:
\begin{equation} C(u_1,\ldots,u_d) = H(u_1,\ldots,u_d) \qquad \forall (u_1,\ldots,u_d) \in \bigtimes_{k=1}^d \Ran F_k. \label{eqn:CS} \tag{$\star\hspace{-1mm}\star\hspace{-1mm}\star$}\end{equation}
Then (\ref{eqn:Sklar}) follows immediately from (\ref{eqn:Sklarsub}) and (\ref{eqn:CS}). Evidently, the values taken by any such copula $C$ outside $\bigtimes_{k=1}^d \Ran F_k$ are totally irrelevant, as they do not even appear in (\ref{eqn:Sklar}). All in all, {\it Th\'eor\`eme} \ref{thm:Sklar2} is akin to part $a)(i)$ of Theorem \ref{thm:pretendSklar}, while clearly {\it Th\'eoreme} \ref{thm:Sklar3} is its part $b)$.

\ppn What about part $a)(ii)$? \cite{Sklar59} does not make any specific mention of the uniqueness of the copula in the continuous case. Rather the contrary, {\it Th\'eor\`eme} \ref{thm:Sklar2} is stated {in general} with an explicit note about the non-uniqueness of the copula. Naturally, for a continuous univariate distribution $F_k$, $\Ran F_k \equiv [0,1]$, thus if each $F_k$ ($k=1,\ldots,d$) is continuous, then $\bigtimes_{k=1}^d \Ran F_k \equiv [0,1]^d$. In that case, (\ref{eqn:CS}) implies that the subcopula is a copula, and since there is no room for arbitrary extension, any copula $C$ satisfying (\ref{eqn:Sklar}) {\it must be} the subcopula, making such $C$ unique. Hence part $a)(ii)$ follows from {\it Th\'eor\`emes} \ref{thm:Sklar1}-\ref{thm:Sklar2} and is not an add-on {\it stricto sensu}, but was apparently not an essential point to make for \cite{Sklar59}.

\ppn This illustrates that Theorem \ref{thm:pretendSklar} should not be regarded as just a concise re-statement of the sequence {\it Th\'eor\`eme 1}, {\it Th\'eor\`eme 2} and {\it Th\'eor\`eme 3}. The substance may be equivalent, but the form is not exactly the same, and this may lead to subtly different reading and interpretation. What is notable is that, although \cite{Sklar59} gives a prominent place to the subcopula -- with {\it Th\'eor\`eme} \ref{thm:Sklar1} explicitly devoted to it -- it has totally disappeared from the `modern' statement Theorem \ref{thm:pretendSklar}, largely consigning it to oblivion. Indeed, in the above classical references, either the subcopula is introduced only in the technical lemmas leading to Theorem \ref{thm:pretendSklar} (\citealp[Lemma 2.3.4]{Nelsen06}; \citealp[Lemma 2.3.3]{Durante15}), or it is not mentioned at all \citep{McNeil05,Joe2015,Hofert18}. Though, it is clear that the only informative part of the copula is the underlying subcopula; and therefore, understanding completely the whys and wherefores of (\ref{eqn:Sklar}) seems conditional on a proper recognition of the role played by $H$. It is my opinion that short-circuiting the subcopula step as in Theorem \ref{thm:pretendSklar} induces overemphasis on the copula(s) $C$ and ultimately unwarranted exploitation of (\ref{eqn:Sklar}) when $C \neq H$. This is, especially, the case when it comes to analysing or modelling dependence, which is the main if not only application of Sklar's theorem.

\ppn Remarkably, Theorem \ref{thm:pretendSklar} does not make any reference to dependence: (\ref{eqn:Sklar}) is merely an analytical result providing an alternative representation of $F_{1\ldots d}$ which may or may not be of any relevance. It is really the {\it interpretation} which we are willing to make of it which brings in the concept of dependence and relates it to copulas. Effectively, it appears from (\ref{eqn:Sklar}) that $C$ is to capture how the marginals $F_1,\ldots,F_d$ interlock inside $F_{1\ldots d}$ -- which is seems fair to called the `dependence structure'. This explains why, early on, copulas were called `{\it dependence functions}'; e.g., in \citet[Definition 5.2.1]{Galambos78} and \citet{Deheuvels79,Deheuvels80}. 

\ppn Though, for playing with the dependence structure of $F_{1\ldots d}$, the subcopula $H$ is the only function worth examining: it always exists, it is always unique, and it always describes unequivocally through (\ref{eqn:Sklarsub}) how to reconstruct $F_{1\ldots d}$ from the marginals $F_1,\ldots,F_d$. Thus, with {\it Th\'eor\`eme} \ref{thm:Sklar1} in hand, it is not clear what is the added value of the copula extension promised by {\it Th\'eor\`eme} \ref{thm:Sklar2} -- that same extension (\ref{eqn:Sklar}) implicitly but exclusively put forward by Theorem \ref{thm:pretendSklar}. Having said this, the fact that the subcopula $H$ contains all necessary information for describing the dependence in $F_{1\ldots d}$ does not imply that it is, in itself, a valid representation of that dependence. Indeed, defined on $\bigtimes_{k=1}^d \Ran F_k$, the subcopula is {in general} not a stand-alone element which can be handled and analysed without reference to marginal distributions, and therefore, cannot isolate a dependence structure as such. In fact, $H$ must adjust to $F_1,\ldots,F_d$ by definition -- again, {\it in general}.

\ppn Now, it so happens that, when all the marginal distributions are continuous, the subcopula takes a very specific form which is invariably a $d$-variate distribution function with continuous uniform margins on $[0,1]$ -- this follows straightforwardly from standard results on functions of random variables applied to (\ref{eqn:Sklarsub}), in particular the Probability Integral Transform (PIT).\footnote{If $X_k$ is a continuous variable with cumulative distribution $F_k$, then $F_k(X_k) \sim \Us_{[0,1]}$ always.} In this case, the subcopula is a copula as per {\it D\'efinition 1}, so $H \equiv C$ (and (\ref{eqn:Sklar}) $\equiv$ (\ref{eqn:Sklarsub})) as observed above, but {\it even more importantly} this (sub)copula is `marginal-distribution-free',\footnote{Where `distribution-free' is taken in the sense of \cite{Kendall53}: {\it free of the parent distribution}. Thus, more specifically here, `marginal-distribution-free', or `margin-free', means {\it free of the marginal distributions of the parent distribution $F_{1\ldots d}$}.} a.k.a.\ `margin-free'. Unbound from any marginal interference, the (sub)copula can now be genuinely understood as capturing the heart of $F_{1\ldots d}$, that is, its dependence structure as such. The representation (\ref{eqn:Sklar}) is then particularly appealing, as it provides an explicit breakdown of a joint distribution into the individual behaviour of the variables of interest (captured by $F_1,\ldots,F_d$) on one hand, and their interdependence structure (captured by $C$) on the other -- with no overlap/redundancy between the two. The entire copula methodology for dependence modelling developed around this neat decomposition and the desirable consequences thereof. 

\ppn It cannot be stressed enough, though, that this pleasant situation only follows as a corollary of two favourable events which occur concurrently {\it when and only when} all the variables involved are continuous: first, the copula appearing in (\ref{eqn:Sklar}) is the subcopula, and second, that subcopula is margin-free. In all other non-continuous situations, any copula $C$ satisfying (\ref{eqn:Sklar}) is nothing more than an arbitrary extension of the subcopula $H$ in (\ref{eqn:Sklarsub}), which itself is {\it not} a satisfactory representation of the dependence of $F_{1 \ldots d}$ as it is not margin-free. The suitability of (\ref{eqn:Sklar}) for analysing and/or modelling dependence becomes then highly questionable. In effect, the validity of any attempt at dependence modelling based on  the typical interpretation of (\ref{eqn:Sklar}) as a clear-cut decomposition `marginals vs.\ dependence', is {\it critically} contingent on the continuity of all the variables.

\ppn Without any reference to the subcopula, the usual statement of `Sklar's theorem' as in Theorem \ref{thm:pretendSklar} does not provide the clues to appreciate that. I questioned above the real benefit of the extension promised by {\it Th\'eor\`eme} \ref{thm:Sklar2} when we have {\it Th\'eor\`eme} \ref{thm:Sklar1}. The question may be rephrased as: why did (\ref{eqn:Sklar}) become the universal baseline, in lieu of (\ref{eqn:Sklarsub})? The only reason I see is that, since a copula is always a distribution supported on $[0,1]^d$ with standard uniform margins by definition, the function $C$ in (\ref{eqn:Sklar}) appears as a standard object in a invariant form (in particular: margin-free); as opposed to the function $H$ in (\ref{eqn:Sklarsub}), whose exact nature is undefined and its specification requiring knowledge of $\bigtimes_{k=1}^d \Ran F_k$. Yet, such invariance of $C$ may only be granted in continuous cases -- but then $H$ enjoys the same desirable property, anyway -- otherwise it is mostly a lure: in fact, the definition of $C$ makes it into a blanket which concealed the fact that, `underneath', its anchor points are fixed by $H$ via (\ref{eqn:CS}). That the gaps between the nodes of $\bigtimes_{k=1}^d \Ran F_k$ may be filled in such a way that $C$ maintains uniform margins is actually little more than an analytical artefact of no obvious relevance when it comes to dependence.

\ppn What adds to the blur is part $a)(ii)$ explicitly contrasting the continuous and non-continuous cases in terms of the (non-)uniqueness of the copula $C$ in (\ref{eqn:Sklar}). This may give the feeling that this is the only notable difference between the two situations, and may consequently divert attention from other questions. Indeed the lack of uniqueness of $C$ and the ensuing problems of {\it model unidentifiability} have often been presented as the main hurdle for practical use of copula methods outside the continuous framework, and have consequently been abundantly commented \citep{Genest07,Trivedi17,Faugeras17,Geenens2020,Nasri23}. In my current view, though, the only consequential difference between continuous and non-continuous cases is that the (sub)copula is margin-free in the former, and not in the latter -- and this seems to have been much less frequently pinpointed as such. 

\ppn What has been discussed is all the `little annoyances' which follow directly from this; e.g., the fact that copula-based dependence measures, such as Kendall's $\tau$ or Spearman's $\rho$, depend on the margins in non-continuous settings (\citealp{Marshall96}; \citealp[Section 4]{Genest07}). Yet, these are only consequences of the lack of margin-free-ness of $C$ which, {\it in itself}, appears to me as the real predicament: in effect we are losing the very reason-of-being of the copula approach, which is precisely its power to dissociate marginal behaviour and dependence structure via (\ref{eqn:Sklar}). For example, \citet[Section 1.6]{Joe2015} motivates resorting to copulas over alternative multivariate models as: ``(...) {\it the copula approach has an advantage of having univariate margins of different types and the dependence structure can be modeled separately from univariate margins}''.  Yet, outside the continuous framework, this alleged separation between dependence structure and margins is clearly violated. All in all, it seems that copula methods applied to non-continuous distributions miss entirely their own point.

\ppn It is, therefore, my opinion that copula-like methods for analysing, modelling and quantifying dependence in non-continuous multivariate distributions should {\it not} be based on (\ref{eqn:Sklar}). I elaborated on this in \cite{Geenens2020}, and proposed an alternative approach for discrete distributions. In a nutshell, the idea is to extract the information about dependence from the subcopula, and to reshape it under the form of a distribution with (discrete) uniform margins -- hence `margin-free' -- this defining a {\it discrete copula}.
	
	\bibliographystyle{../../elsarticle-harv-cond-mod}
	\setlength{\bibsep}{0cm}
	\def\bibfont{\footnotesize} %if class is 11 \footnotesize = 9pt $\small = 10pt http://en.wikibooks.org/wiki/LaTeX/Formatting#Sizing_text
	\bibliography{../../libraries-copula}

\end{document}